\documentclass[twocolumn,aps,showpacs]{revtex4-1}

\usepackage{amsmath}
\usepackage{amssymb}
\usepackage{graphicx}

\usepackage[T1]{fontenc}
\usepackage[english]{babel}
\usepackage{caption}

\makeatletter

\begin{document}

\title{Study of external electron injection and trapping in the blow-out regime}
\author{Jari Pronold}
\email{jari.pronold@hhu.de}
\affiliation{Institut f\"{u}r Theoretische Physik I, Heinrich-Heine-Universit\"{a}t D\"{u}sseldorf, D-40225 Germany}
\author{Johannes Thomas}
\email{johannes.thomas@uni-duesseldorf.de}
\affiliation{Institut f\"{u}r Theoretische Physik I, Heinrich-Heine-Universit\"{a}t D\"{u}sseldorf, D-40225 Germany}
\author{Alexander Pukhov}
%\email{pukhov@tp1.uni-duesseldorf.de}
\affiliation{Institut f\"{u}r Theoretische Physik I, Heinrich-Heine-Universit\"{a}t D\"{u}sseldorf, D-40225 Germany}
%\date{\today}
%========================================================================================================================================
%========================================================================================================================================
%========================================================================================================================================
\begin{abstract}
In this work we study electron side-injection and trapping in the blow-out regime in deep plasma channels. We analyze the maximum angle of injection, for which at least 90\% of the injected electrons are trapped. We discuss the dependence of this angle on the electrons' initial energy and their injection positions. In the scope of a semi-analytical blow-out model we show that the injection position is a less critical factor for trapping if electrons are injected into deep plasma channels. PIC simulations and analytical approximations support our results from the semi-analytical model.
\end{abstract}
\maketitle
%========================================================================================================================================
%========================================================================================================================================
%========================================================================================================================================
\section{Introduction}
\label{sec:intro}
Plasma-based electron acceleration methods take advantage of wakefield excitation by either a relativistic electron bunch for plasma wakefield acceleration (PWFA) \cite{Chen1985, Rosenzweig1988} or an intense laser pulse for laser wakefield acceleration (LWFA) \cite{Tajima1979}. In both cases the longitudinal electric wakefield gradient is in the order of 100 GV/m, which is orders of magnitudes higher than in conventional accelerators \cite{Kostyukov2015, Esarey2009}. If the laser pulse intensity reaches a certain threshold value, the wakefield breaks and a solitary electronic cavity, called the bubble, is formed \cite{Pukhov2002, Jansen2014, Lu2007}. It is a nearly spherical region with uniform accelerating fields that propagates with almost speed of light c \cite{Kostyukov2004}. A similar structure can be created, if a dense relativistic electron beam excites a so called "blow-out"\cite{Rosenzweig1991, Lotov2004}. In both cases the expelled electrons gather in a thin sheath on the border of the cavity, while those electrons, which become trapped inside the wakefield, form a dense witness bunch - the so called beam load.

The major feature, that characterizes the bubble and the blow-out regime, is the quasi-monoenergetic energy spectrum of the fast electrons inside the beam load. However, if the total charge of the beam load exceeds a certain threshold, the plasma cavity
structure is reshaped and the effective accelerating field is modified. This in turn affects final beam properties like maximum energy, energy spread and transverse emittance \cite{Golovanov2016b, Lu2007, Tzoufras2009, Couperus2017}. As a consequence, it is necessary to find a beam loading technique which gives maximum control over injection parameters like total charge, initial momenta and initial positions inside the wakefield. An especially effective method is the lateral or on-axis injection of pre-accelerated electron bunches.

When using the on-axis injection technique, the driver and the electron bunch propagate on the same axis. In the case of an intense laser driver this may lead to bunch scattering if the laser pulse passes through the electron bunch in vacuum. The resulting limitation of the number of trapped particles then would lower the quality of the accelerated bunch. To bring this problem under control, the side injection method has been proposed \cite{Luttikhof2009} and applied to proton-driven wakefields \cite{Pukhov2011} already.

In this work, we consider side injection of pre-accelerated electron bunches into a blow-out at a small angle $\vartheta$ (see Fig.\ref{pic:3d}). We study the dependence of the critical injection angle $\vartheta_\text{crit}$, for which at least 90\% of the injected particles are trapped, on the injection position, the initial electron energy and the radial plasma density profile in a deep channel. Our work is done in the scope of a semi-analytical blow-out model and compared to analytical approximations and particle-in-cell simulations. We show that external injection into blow-outs is less critical in deep channels than in homogeneous plasma. A comparison of our results from analytical predictions to test particle simulations in a quasi-static blow-out model and to PIC simulations indicates that in homogeneous plasma it is favorable to inject electron bunches on-axis while in channeled plasma it is possible to trap bunches which have been injected off-axis. However, this advantage is compensated by the need for a higher initial focussing of the injected electron beam.

In section \ref{sec:modell} we present the semi-analytical model and derive the equations which are solved numerically in section \ref{sec:simu}. In \ref{sec:ana} we derive an analytical description of $\vartheta_\text{crit}$ which is compared to test particle simulations in section \ref{sec:simu}. A further comparison to simulations with the fully electromagnetic version of the three-dimensional PIC code VLPL \cite{Pukhov1999,Pukhov2016} is given in section \ref{sec:compare}.
\\

%========================================================================================================================================
%========================================================================================================================================
%========================================================================================================================================

\section{The analytical model}
\label{sec:modell}

\begin{figure}[t]
	\centering
	\includegraphics[width=0.47\textwidth]{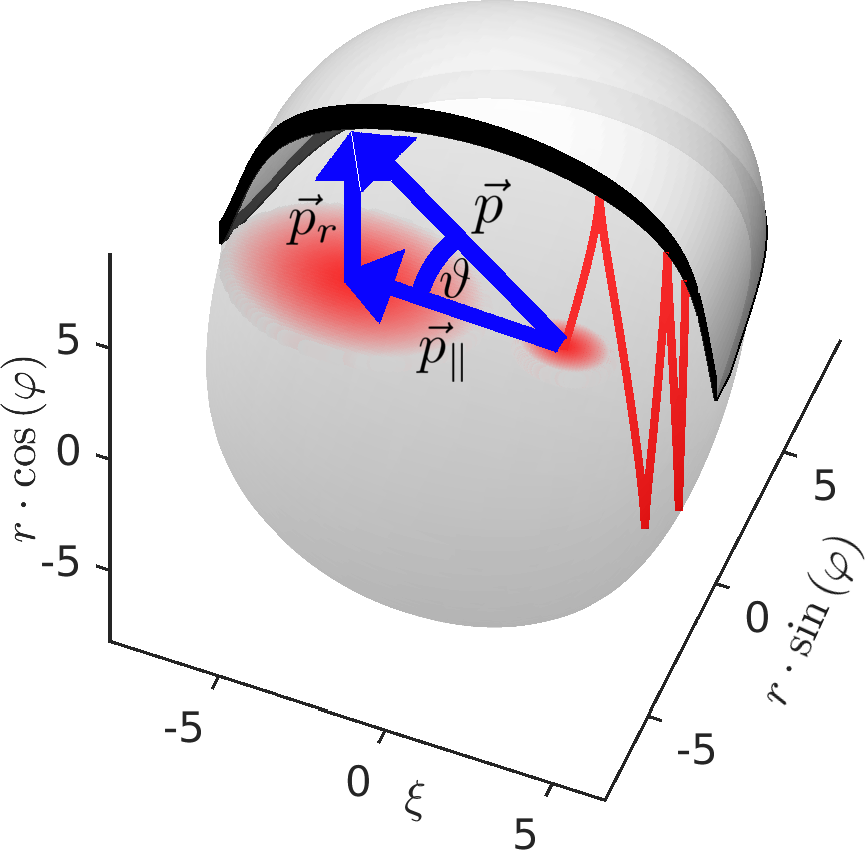}
	\caption{Semi-analytic model of a a radially inhomogeneous $s_i(r) \propto \alpha r^2$  blow-out. On the left is the driving bunch, on the right the injected bunch. The trajectory of a test particle, red solid line, and its injection angle $\tan(\theta) = p_r/p_\parallel$, in blue, are shown.}
	\label{pic:3d}
\end{figure}

Besides analytical and semi-analytical bubble and blow-out models for homogeneous plasma \cite{Kostyukov2009, Kostyukov2010, Lu2006a, Yi2011, Yi2013, Pak2010, Zeng2012, Thomas2014}, there are also more general models for channeled plasmas \cite{Thomas2016, Golovanov2017b, Golovanov2016b, Golovanov2016c} describing the blow-out envelope and the fields in terms of the radial distance to the symmetry axis $r_b$ in a moving frame of reference. In this frame all fields and sources are quasi-static, which means that they depend on $\xi=ct-z$ solely. In the following we normalize coordinates to the inverse electron wave number $k_p^{-1}=c/\omega_p$, velocities to the speed of light $c$, fields to $E_0 = m_ec\omega_p/e$ and time to the electron plasma frequency $\omega_p^{-1} = \sqrt{m_e/(4\pi e^2 n_0)}$, where $m_e$ is the electron mass and $n_0$ is a certain density in the system to which the electron density $n_e(r)$ and the ion density $\rho_{ion}$ are normalized. After this transformation we further take the cylindrical symmetry of the system into account and write the fields within a blow-out ($r\leq r_b$) for the case of a large blow-out \cite{Thomas2016}
\begin{align}
	B_\varphi = \frac{ S_{Ib}}{2}\frac{r}{r_b}\left( \frac{r_b^{\prime 2}}{r_b} - r_b^{\prime\prime} \right) - \frac{1}{2} s_i(r_b) r r_b^{\prime 2} - \frac{\Lambda(\xi)}{r} 
	\\
	E_r =  B_\varphi - \frac{S_I(r)}{r},
	\qquad
	E_z =  - S_{Ib} \frac{r_b^{\prime}}{r_b} 
	\label{eq:EundB}.
\end{align}
Here, $s_{ib} = s_i(r_b)$ and $S_{Ib} = S_I(r_b)$ are abbreviations for the negative ion density $s_i(r) = -\rho_\text{ion}(r)$ and its weighted integral $S_I(r) = \int_0^rs_i(r^\prime)r^\prime \, dr^\prime$. The function $r_b(\xi)$ can be calculated from the differential equation
\begin{align}
&	S_{Ib}r_br_b'' + s_{ib}r_b^2r_b'^2 + S_{Ib} = -2\Lambda(\xi), \label{ODE}
\end{align}
where $\Lambda(\xi) = -\int_0^{r_b}J_z(\xi,r')r'dr'$ is the weighted integral of the longitudinal current density created by
electron bunches within the blow-out. In figure \ref{pic:3d} the solution to this equation is shown for $\rho_{ion} \propto r^2$. The red line is the trajectory of a test particle starting with initial momentum $\vec{p}_0=p_{r}\vec{e}_r +p_{||}\vec{e}_z$ inside the witness bunch (small red cloud). The surrounding electron sheath is indicated by a black layer.
In this figure the driving electron bunch (large red cloud) moves along the negative $\xi$-axis.

The ODE (\ref{ODE}) is valid in the relativistic approximation $r_b/(S_{Ib}M_1(0))\ll\Delta\ll r_b$, where $\Delta$ is the thickness of the sheath and $M_0(x) = \int_x^\infty g(y) \,dy$ is the zeroth moment of an arbitrary function $g(y)$ describing the shape of the boundary \cite{Golovanov2016}. In our present work we follow \cite{Lu2006, Thomas2016}, where $g(y) = \Theta(1-y)$ was used to model a rectangular shaped layer profile by the Heaviside step-function $\Theta(x)$. With this model for the electron layer we follow the argumentation of Golovanov et al. \cite{Golovanov2017b} and strongly simplify the fields inside the sheath by a lowest-order perturbation theory with respect to $\epsilon$.

In general, the first order wakefield potential for $r>r_b$ can be expressed as
\begin{align}
\Psi(\xi,r) = -S_{Ib}\epsilon\int_{R(r,r_b)}^\infty M_0(Y) \, dY,
\end{align}
where $R(r,r_b) = (r-r_b)/\Delta$. Since $g(y) = \Theta(1-y)$, we find
\begin{align}
\Psi(\xi,r) = & -\frac{S_{Ib}}{2\epsilon} \left(\zeta-\epsilon\right)^2,
\end{align}
with $\zeta = (r-r_b)/r_b$ and thus
\begin{align}
E_z= \frac{\partial \Psi}{\partial \xi}= -\frac{\epsilon - \zeta}{\epsilon}\frac{S_{Ib}r_b^\prime}{r_b},
 \label{eq:ez_rand}
\end{align}
for $r>r_b$. This result is equivalent to Golovanov et al. \cite{Golovanov2017b}, where an exponentially decaying sheath source was assumed. The other two field components are $B_\varphi = - \partial_\xi A_r - \partial_r A_z$ and $E_r = - \partial_r \Psi + B_\varphi$, where
\begin{align}
\frac{\partial A_r}{\partial \xi} = \frac{1}{r} \int_0^r \frac{S_{Ib}} {\epsilon} \frac{r_b^{\prime 2}}{r_b^2} r^\prime  dr^\prime
\label{darxi}
\end{align}
and 
\begin{align}
\frac{\partial A_z}{\partial r} = - \frac{1}{r} \int_0^r J_z(\xi,r^\prime)r^\prime dr^\prime.
\label{dazr}
\end{align}
To calculate the plasma return current $J_s(\xi)$ we follow the sheath model presented in \cite{Thomas2016} and see that it is connected to the wakefield potential by Ampere's circuit law
\begin{align}
\oint_{\partial P} \vec{B} \cdot d\vec{l} = \int_P \left( \vec{J} + \frac{d\vec{E}}{dt} \right) \cdot d\vec{s}.
\end{align}
If we decide that the surface $P$ is the transverse plane with the unit normal vector $\hat{n} = \vec{e}_z$, integrating over the whole plane gives
\begin{align}
\int_{\mathbb{R}^2} \left( J_z + \frac{\partial E_z}{\partial t} \right) dx \, dy = 0.
\end{align}
We know that $\Psi$ and $J_z$ vanish outside the surface and we know that $\int_0^{r_b} J_z r \, dr = - \Lambda (\xi)$ connects the sheath current to the wakefield potential. Thus 
\begin{align}
\int_0^{r_b+\Delta} J_z r \, dr = - \int_0^{r_b+\Delta} \frac{\partial^2 \Psi}{\partial \xi^2} r \, dr
\end{align}
and the plasma return current can be written as
\begin{align}
	J_s(\xi) = & \frac{-2}{(r_b+\Delta)^2 - r_b^2} \left( \int_0^{r_b+\Delta} \frac{\partial^2 \Psi}{\partial \xi^2} r \, dr -\Lambda(\xi) \right).
	\label{eq:js_jo}
\end{align}
Using the Lorenz gauge, the normalized Poisson equations and equation \eqref{eq:js_jo} for $r_b < r < r_b + \Delta$, we get
\begin{align}
r B_\varphi = - \frac{X^2 - (1+\zeta)^2}{ X^2 - 1} \left( \int_0^{r_b} \frac{\partial^2 \Psi}{\partial \xi^2}r \, dr - \Lambda(\xi) \right)
\nonumber
\\
+ \int_{r_b}^r\frac{\partial^2\Psi}{\partial \xi^2}r \, dr + \frac{1 - (1+\zeta)^2}{X^2 - 1} \int_{r_b}^{r_b+\Delta} \frac{\partial^2\Psi}{\partial \xi^2}r \, dr.
\label{eq:lange_b_phi}
\end{align}
At our level of precision
\begin{align}
\frac{X^2 - (1+\zeta)^2}{ X^2 - 1} \approx & \frac{\epsilon - \zeta}{\epsilon},
\quad
\frac{\partial^2 \Psi}{\partial \xi^2} \approx \frac{S_{Ib}}{\epsilon}\frac{r_b^{\prime 2}}{r_b^2}
\end{align}
and the last two terms in equation \eqref{eq:lange_b_phi} vanish so that
\begin{align}
B_\varphi =  \frac{\zeta - \epsilon}{\epsilon} \left(\frac{S_{Ib}}{2 \epsilon} r_b^{\prime 2} + \Lambda(\xi)\right)\frac{1}{r}. \label{Bphi}
\end{align}
Finally, with $\partial_r \Psi = -(\zeta-\epsilon)S_{Ib}/(\epsilon  r_b)$, the radial component of the electric field can be expressed as
\begin{align}
E_r = B_\varphi + \frac{\zeta - \epsilon}{\epsilon} \frac{S_{Ib}}{r_b}. \label{Er}
\end{align}

The fields in (\ref{eq:EundB}) for $r\leq r_b$ and the fields in (\ref{eq:ez_rand}), (\ref{Bphi}) and (\ref{Er}) for $r>r_b$ are the base for our blow-out model. In section \ref{sec:simu} we will solve the equations of motion of a test particle in these fields to find the maximum injection angle in dependence of the particle's position, the particle's initial energy, the blow-out radius and the plasma density profile. In the following section we derive an approximation for the maximum injection angle assuming a spherical blow-out shape. A comparison between predictions of this approximation to simulations from section \ref{sec:simu} and PIC simulations is given in section \ref{sec:compare}.

%========================================================================================================================================
%========================================================================================================================================
%========================================================================================================================================

\section{Estimation of the maximum injection angle}
\label{sec:ana}
In the following we estimate a trapping condition for test particles with initial radial momentum $p_{r,0}$, initial parallel momentum $p_{||,0}\gg p_{r,0}\gg1$, and initial position $(\xi_0,0)$ in terms of the angle of injection $\vartheta_\text{crit} =\arctan(p_{r,0} / p_{||,0})$. We derive a closed formula for the critical angle $\vartheta_\text{crit}=\max\vartheta$ which barely allows for trapping of a test electron.
 
For simplicity we assume that trapping occurs during a characteristic time which is approximately the first quarter oscillation. In this time a particle reaches its maximal distance to the symmetry axis $r_\text{max}$ and has traveled a certain distance $\Delta\xi$ on the $\xi$-axis. In our analysis we assume that the blow-out is a perfect circle with radius $R$ so that the simplest trapping condition is 
\begin{align}
& r_\text{max}\leq \sqrt{R^2 -\xi_\text{max}^2}, && \xi_\text{max}=\xi_0+\Delta\xi \label{condition}.
\end{align}
The kinetic energy of the particle changes adiabatically slow during the first quarter oscillation. Thus we conclude that radial momentum is completely transferred into parallel momentum during the same time. 

To calculate $r_\text{max}$ we solve the equations of motion for a test particle within the blow-out. For a circular shaped electron sheath, in cylindrical coordinates and in a co-moving frame of reference they are
\begin{align}
	\dot{\textbf{p}} = - \left( \textbf{E} + \dot{\textbf{r}}_\bot \times \textbf{B} \right), && \dot{\textbf{r}} = \frac{\textbf{p}_\bot}{\gamma} + \vec{e_z} \left( 1 -\frac{p_{||}}{\gamma} \right) \label{motion}
\end{align}
with $E_z=\xi/2$ and $E_r=-B_\varphi=r/4$ (also compare \cite{Kostyukov2004, Kostyukov2009}). Solving the radial motion for $p_\parallel/\gamma \approx 1$ we find  $\ddot{r} = - r/(2\gamma)$, which describes the betatron motion of an electron with frequency $\omega=1/\sqrt{2\gamma}$. Using the boundary conditions $r(t=0) = 0$ and $\dot{r}(t=0) = p_{r,0}/\gamma$ the betatron motion in the first quarter can be approximated by
\begin{align}
  r(t) = \sqrt{\frac{2}{\gamma}} p_{r,0} \sin{\left( \frac{t}{\sqrt{2\gamma}}\right)},
  \label{eq:r}
\end{align}
which shows that the test particle reaches its maximum distance to the symmetry axis $r_\text{max} = \sqrt{2/\gamma}\, p_{r,0}$ at time $t_\text{max} = \sqrt{\gamma/2}\,\pi$. 

\begin{figure}[t]
	\centering
	\includegraphics[width=0.47\textwidth]{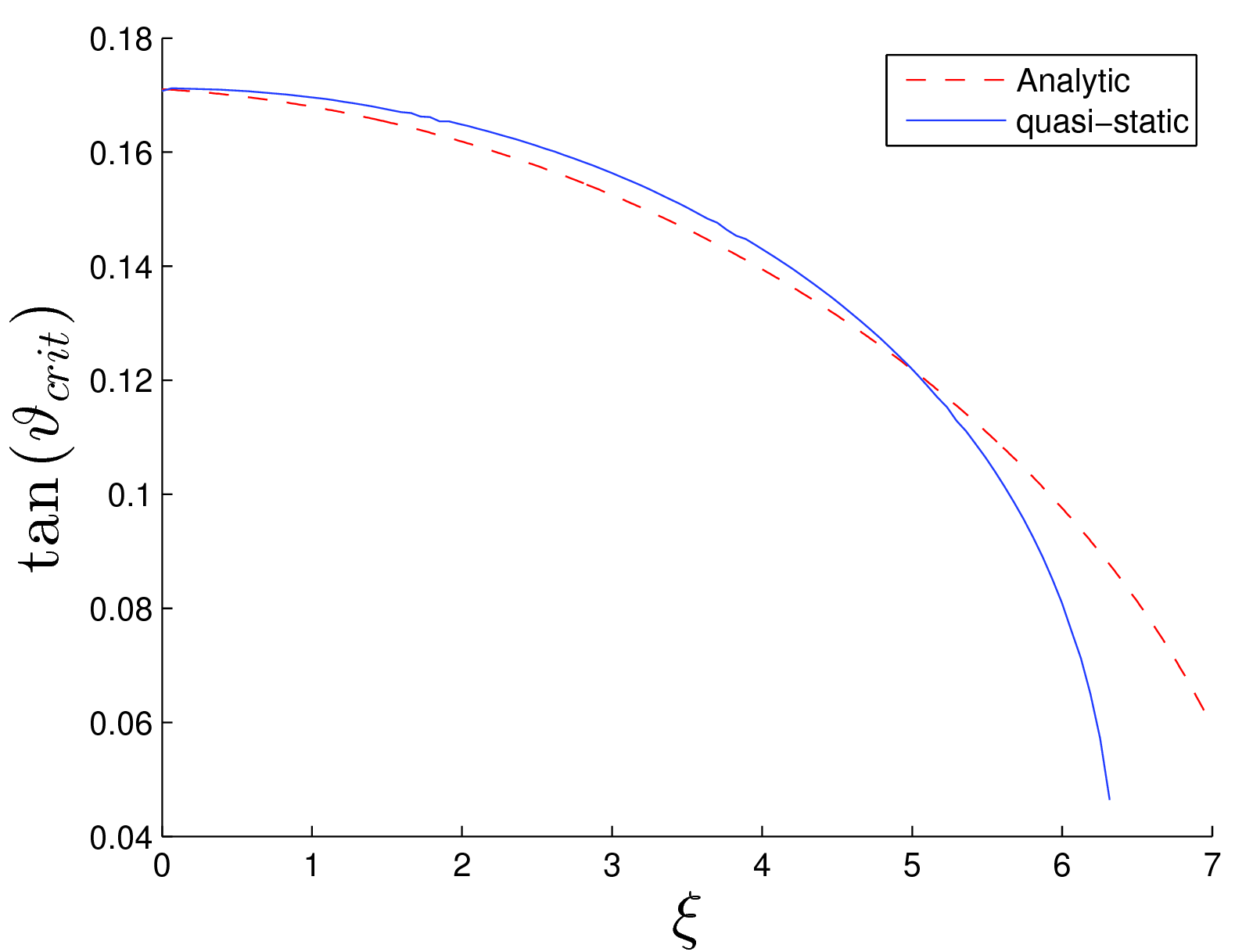}
	\caption{Analytical solution (red dotted line) and numerical solution (blue line) of the critical angle of injection $\tan{(\vartheta_\text{crit})}$ in dependence of the initial position $(\xi_0,0)$ of a test electron inside a blow-out.}
	\label{pic:ana_quasi_static_angle}
\end{figure}

To calculate how far the particle has moved on the $\xi$-axis during the time $t_\text{max}$, we solve the equations of motion for $\xi$. Since $p_{||,0}\gg p_{r,0}\gg1$ and $\gamma$ changes adiabatically slow during the first quarter of the first betatron oscillation, the equations of motion (\ref{motion}) can be combined to 
\begin{align}
  \dot{\xi} & = 1 - \sqrt{ 1 - \dot{r}^2}.
\end{align}
Eq.\eqref{eq:r} and the boundary condition $\xi(t=0)=\xi_0$ then yield
\begin{align}
  \xi(t) & = t - \sqrt{\frac{2}{\gamma}}p_\parallel\int_0^{t/\sqrt{2\gamma}} dt^{\prime}  \sqrt{ 1 + \frac{p_{r,0}^2}{ p_{\parallel,0}^2 } \sin^2{\left( t^{\prime}\right)} } \nonumber
  \\
  & = t - \sqrt{\frac{2}{\gamma}}p_\parallel E\left(\frac{t}{\sqrt{2 \gamma}} \middle| - \frac{p_{r,0}^2}{ p_{\parallel,0}^2 }\right) + \xi_0, \label{xi}
\end{align}
where 
\begin{align}
	E(\varphi|m) = 	\int_0^\varphi \sqrt{1-m\sin^2(\vartheta)}d\vartheta \label{incomplete}
\end{align}
is the incomplete elliptic integral of the second kind. If we substitute $t=t_\text{max}$ in Eq.(\ref{xi}) $E$ becomes a complete elliptic integral and we can make use of the relation $E(-m)=\sqrt{1+m}E(m/(1+m))$ so that
\begin{align}
 \xi_\text{max} = \sqrt{\frac{\gamma}{2}}\pi - \sqrt{2\gamma} E\left(\frac{ p_{r,0}^2}{\gamma^2}\right)  + \xi_0
	\label{eq:xi_max_komplett}
\end{align}
with
\begin{align}
E(m) = 	\frac{\pi}{2}\left[ 1-\sum_{n=1}^\infty\left(\frac{(2n-1)!!}{(2n)!!}\right)^2 \frac{m^n}{2n-1} \right].
\end{align}
If we expand $E(p_{r,0}^2/\gamma^2)$ for $p_{r,0}^2/\gamma^2\approx 0$ up to the second non-vanishing order, Eq.\eqref{eq:xi_max_komplett} becomes
\begin{align}
 \xi_\text{max} \approx \frac{\pi}{4}\frac{p_{r,0}^2}{\sqrt{2\gamma^3}} + \xi_0
 \label{eq:xi(t_max)}
\end{align}
and a formula for the initial radial momentum $p_{r,0}$ can be calculated from the trapping condition (\ref{condition}). Since we assumed that $p_{r,0}\ll p_{||,0}$, we take only terms up to second order in $p_{r,0}/\gamma$ into account so that
\begin{align}
	p_{r,0}^2 \leq \frac{R^2 - \xi_0^2 }{\left( 4\sqrt{2\gamma} +\pi\xi_0\right)}\sqrt{2\gamma}^3 .
	\label{eq:pr0quadrat}
\end{align}
With Eq.(\ref{eq:pr0quadrat}) it is possible to express the critical angle in terms of the initial position of a test electron inside a spherical blow-out as
\begin{align}
\tan{(\vartheta_\text{crit})} = \left( \frac{\left( 4\sqrt{2\gamma} + \pi\xi_0\right) \sqrt{\gamma}} {(R^2 - \xi_0^2)  \sqrt{2}^3} -1\right)^{-1/2}.
\label{eq:tancrit}
\end{align}
In figure \ref{pic:ana_quasi_static_angle} we compare this strongly simplified trapping condition to numerical solutions of the equations of motion (\ref{motion}) for a blow-out with an electron sheath calculated from Eq.(\ref{ODE}). For our simulations we  assumed the fields (\ref{eq:EundB}) for $r\leq r_b$, as well as (\ref{eq:ez_rand}), (\ref{Bphi}) and (\ref{Er}) for $r>r_b$. 
The initial positions of the test electrons are located on axis ($r_0=0$) in the accelerating blow-out phase. As figure \ref{pic:ana_quasi_static_angle} shows, both results are in good agreement as long as the blow-out can be approximated by a sphere i.e. near the blow-out center. In the rear part, i.e. for large $\xi$, strong deviations occur due to the non-spherical form of the electron sheath.

%========================================================================================================================================
%========================================================================================================================================
%========================================================================================================================================

\section{Simulations of test particles}
\label{sec:simu}
In the end of the last chapter we compared an analytical estimation of the critical injection angle $\vartheta_\text{crit}$ to numerical simulations for the special case of electrons which are initially located on the symmetry axis of the blow-out. In this section we study the dependence of $\vartheta_\text{crit}$ on the initial positions in the whole blow-out, on the initial electron energy and on the plasma density profile systematically. We discuss results from numerical simulations solving the equations of motion (\ref{motion}) for a blow-out with an electron sheath calculated from Eq.(\ref{ODE}). For our simulations we consider the blow-out model introduced in section \ref{sec:modell}, where the fields in (\ref{eq:EundB}) belong to the inner wakefield ($r\leq r_b$) and those in (\ref{eq:ez_rand}), (\ref{Bphi}) and (\ref{Er}) determine the electron motion inside the surrounding sheath with thickness $\Delta$ which is approximately 1\% of the maximum blow-out radius $r_{b,\text{max}}$. To calculate the integral current $\Lambda(\xi)$ we assumed a cylinder symmetric electron driver with parabolic density profile
\begin{align}
	n_b = n_{b,0} \left(1-\left(\frac{2r}{\sigma_r}\right)^2 -\left(\frac{2(\xi-\xi_d)}{\sigma_\xi}\right)^2\right)
	\label{eq:flat_top}
\end{align}
%and $n_{b,0} = 8.8$, $\sigma_r = 4.8k_p^{-1} = 0.8\lambda_p$, as well as $ \sigma_\xi = 7.2k_p^{-1} = 1.2\lambda_p$. The shift $\xi_d$ is chosen such that $r_b(\xi=0)=r_{b,\text{max}}$. The ion density of the background plasma is modeled in polynomial form
and $n_{b,0} = 8.8$, $\sigma_r = 4.8$, as well as $ \sigma_\xi = 7.2$. The shift $\xi_d$ is chosen such that $r_b(\xi=0)=r_{b,\text{max}}$. The ion density of the background plasma is modeled in polynomial form
\begin{align}
	\rho_\text{ion}(r) & = \alpha r^n,
\end{align}
where $\alpha$ is chosen such that the plasma density at distance $r_{b,max}$ from the symmetry axis is approximately $1$. 

\begin{figure*}
\centering
\includegraphics[width=\textwidth]{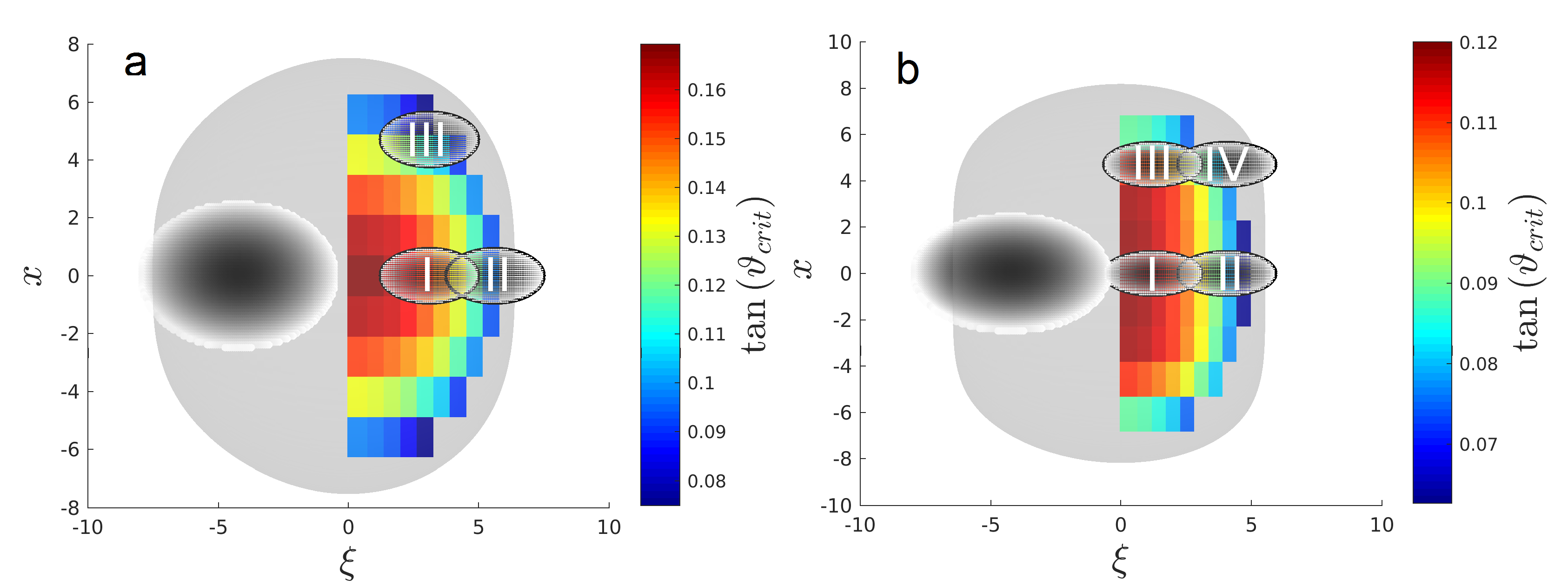}
\caption{Blow-out (gray sphere) driven by a dense particle bunch (ellipse on the left-hand side) and subdivision of the accelerating phase into boxes for a) homogeneous plasma and b) channeled plasma with $\rho_\text{ion}\propto r^2$. The color of a box represents the critical angle of injection for which at least 90\% of all electrons could be trapped. Numbered ellipses on the right-hand side mark injection positions of pre-accelerated electron bunches we simulated in PIC.}
\label{pic:winkel}
\end{figure*}

For our simulations we subdivide the accelerating part of the blow-out into small boxes (see Fig.\ref{pic:winkel}a and Fig.\ref{pic:winkel}b), each containing 2000 non-interacting test electrons. The largest angle which allows for trapping of at least 90 \% of the particles inside a box defines the critical injection angle for that particular box and can be compared to the analytical prediction from Eq.(\ref{eq:tancrit}). In figures \ref{pic:winkel}a and \ref{pic:winkel}b the box colors indicate the value of $\vartheta_\text{crit}$ which is high in the wakefield center (red boxes) and low near the border (blue boxes). The numbered ellipses mark injection positions of bunches we simulated with PIC. They are discussed in the following section.

For homogeneous plasma ($\rho_\text{ion}=1$) Fig.\ref{pic:winkel}a shows that the inner critical injection angles are twice as large as the outer ones but are still in a linear regime, i.e. in a range where  $\tan(\vartheta_\text{crit})\approx \vartheta_\text{crit}$. Since $\vartheta_\text{crit}$ is maximal in the center of the blow-out and declines radially, it is favorable to inject pre-accelerated electron bunches on-axis into blow-outs in homogeneous plasma. For a blow-out in a channeled plasma with $\rho_\text{ion}\propto r^2$ Fig.\ref{pic:winkel}b shows a slightly different result. As can be seen $\vartheta_\text{crit}$ reaches its maximum in a much wider range in the central part of the wake. As a consequence, the distance to the symmetry axis is less important for trapping than in homogeneous plasmas. However, the color scaling in Fig.\ref{pic:winkel}a and Fig.\ref{pic:winkel}b shows that the maximum of $\vartheta_\text{crit}$ is 25\% less for channeled plasmas then for homogeneous plasmas. This is important because the possibility to trap bunches which have been injected off-axis in channeled plasma is compensated by the need for a higher initial focussing.

In the next section we compare results from PIC simulations to the analytical predictions from section \ref{sec:ana} and the results from test particle simulations presented in this section.

\section{Comparison to PIC simulations}
\label{sec:compare}
\begin{figure}[b]
	\centering
	\includegraphics[width=0.47\textwidth]{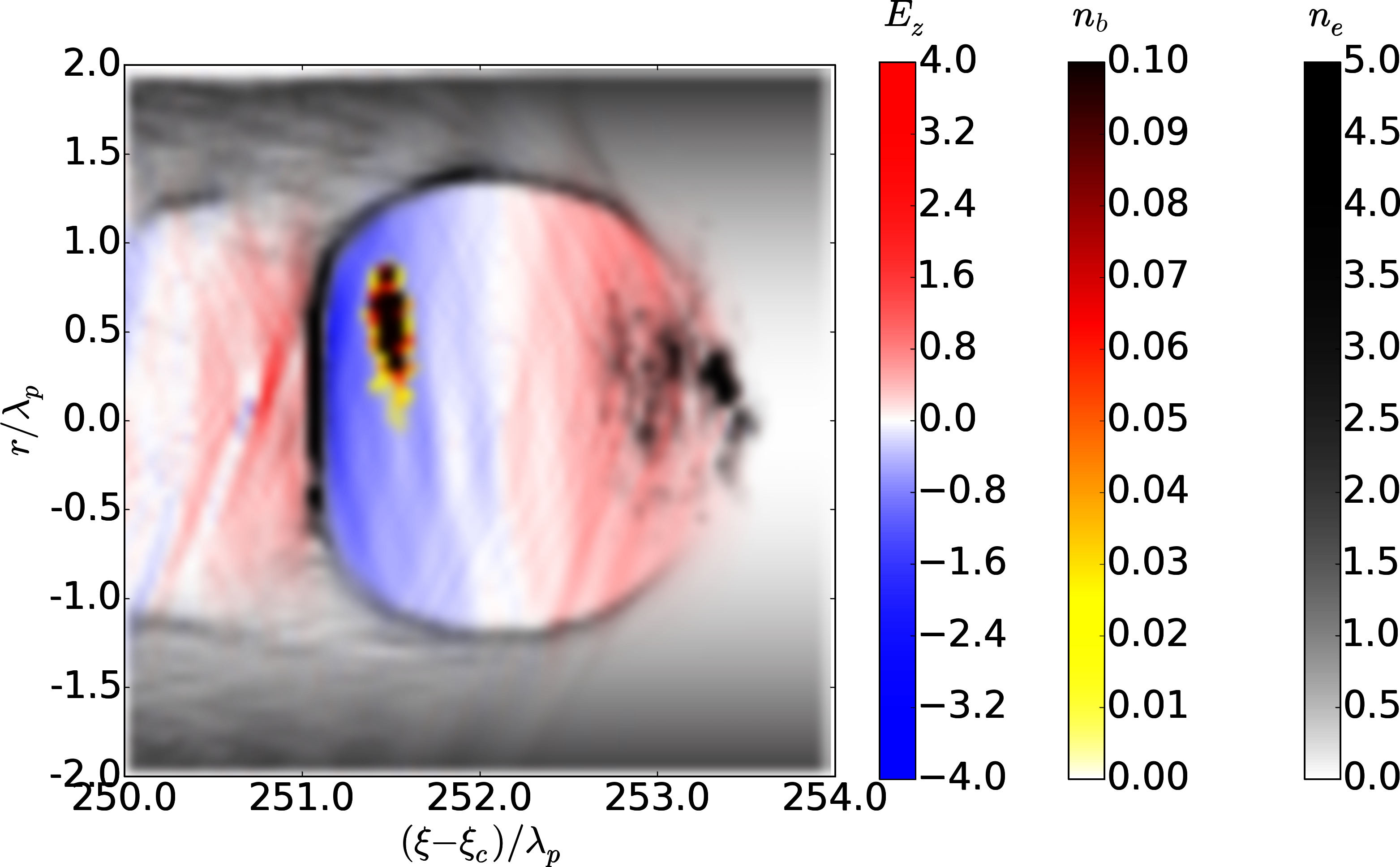}
	\caption{PIC simulation of a lateral injected electron beam (yellow framed bunch) into a blow-out in a parabolic $\rho_\text{ion}\propto r^2$ plasma density channel. The longitudinal electric field $E_z$ accelerates the trapped electrons $n_b$. The wakefield is excited by a dense electron bunch (high dense electrons on the right hand side) in the background electron density $n_e$.}
	\label{pic:pic_quadratic}
\end{figure}
Our PIC simulations are carried out using the fully electromagnetic version of the three-dimensional PIC code VLPL \cite{Pukhov2016, Pukhov1999}. An exemplary simulation in a channeled plasma with parabolic density profile $\rho_\text{ion} \propto r^2$ is shown in Fig.\ref{pic:pic_quadratic}. Here, the longitudinal electric field $E_z$, the density of the injected electron bunch $n_b$ and the electron plasma density $n_e$ are shown. Both the injected and the driving bunch are modeled by the parabolic density profile \eqref{eq:flat_top}. For the driver $n_{b,0}$ is chosen such that its total charge is $1.7 \text{nC}$ for $\sigma_r = 0.8 \lambda_p$ and $ \sigma_\xi = 1.2 \lambda_p$. The injected electron bunch has a lower total charge of $10 \text{pC}$ and a smaller spatial extension of $\sigma_r = 0.3 \lambda_p$ and $\sigma_\xi = 0.6 \lambda_p$. The electrons inside the driver have an energy of $E_d = 5$ GeV each, while the witness bunch consists of pre-accelerated electrons with an initial energy of $E_w = 500$ MeV. The injection angle $\tan(\vartheta)=p_r/p_{||}\approx 0.08$ is large enough to cause large amplitude betatron oscillations of the injected beam and is small enough to trap 99.8 \% of the pre-accelerated electrons. 
\begin{figure*}
	\centering
	\includegraphics[width=0.47\textwidth]{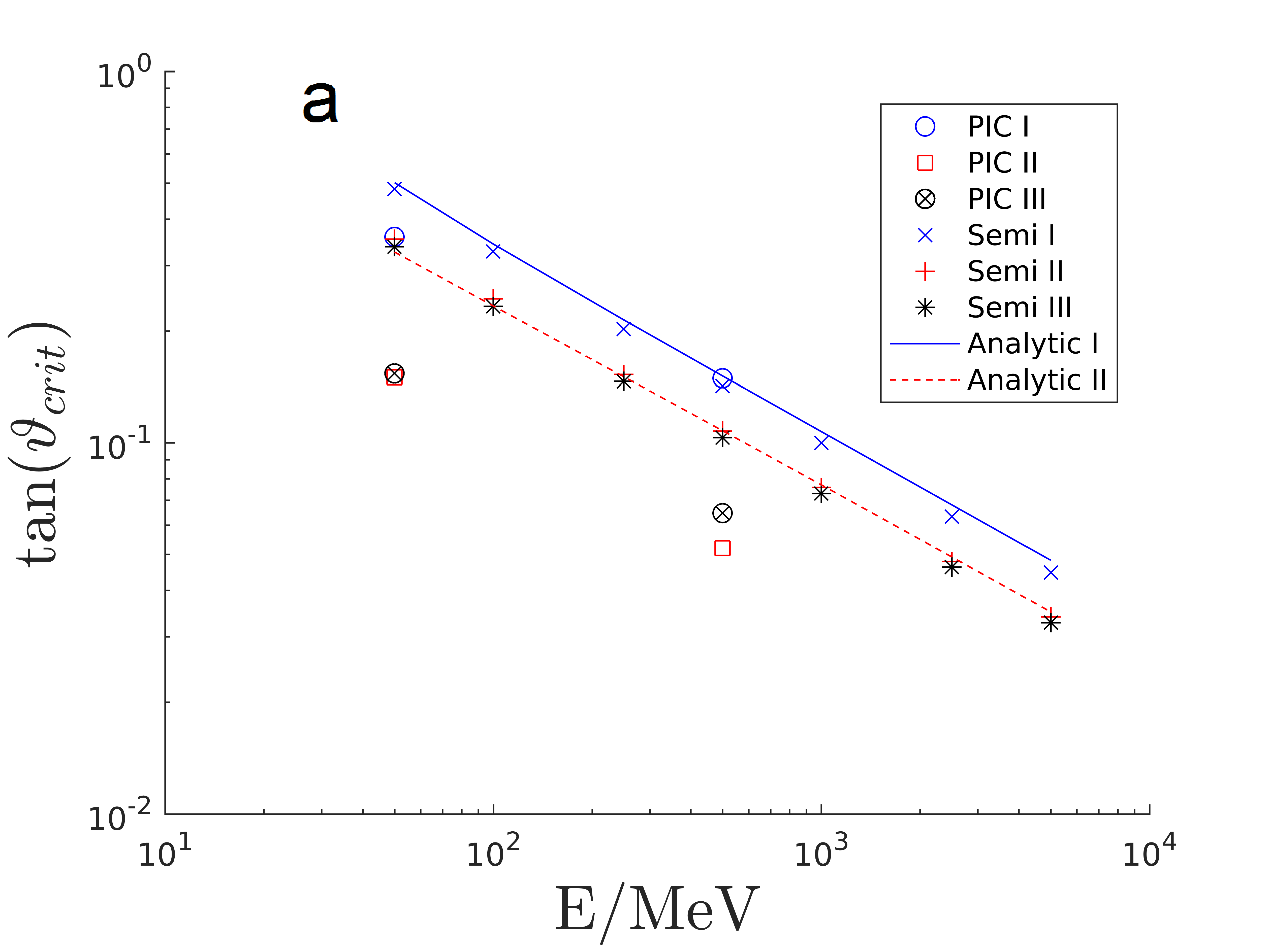}
	\includegraphics[width=0.47\textwidth]{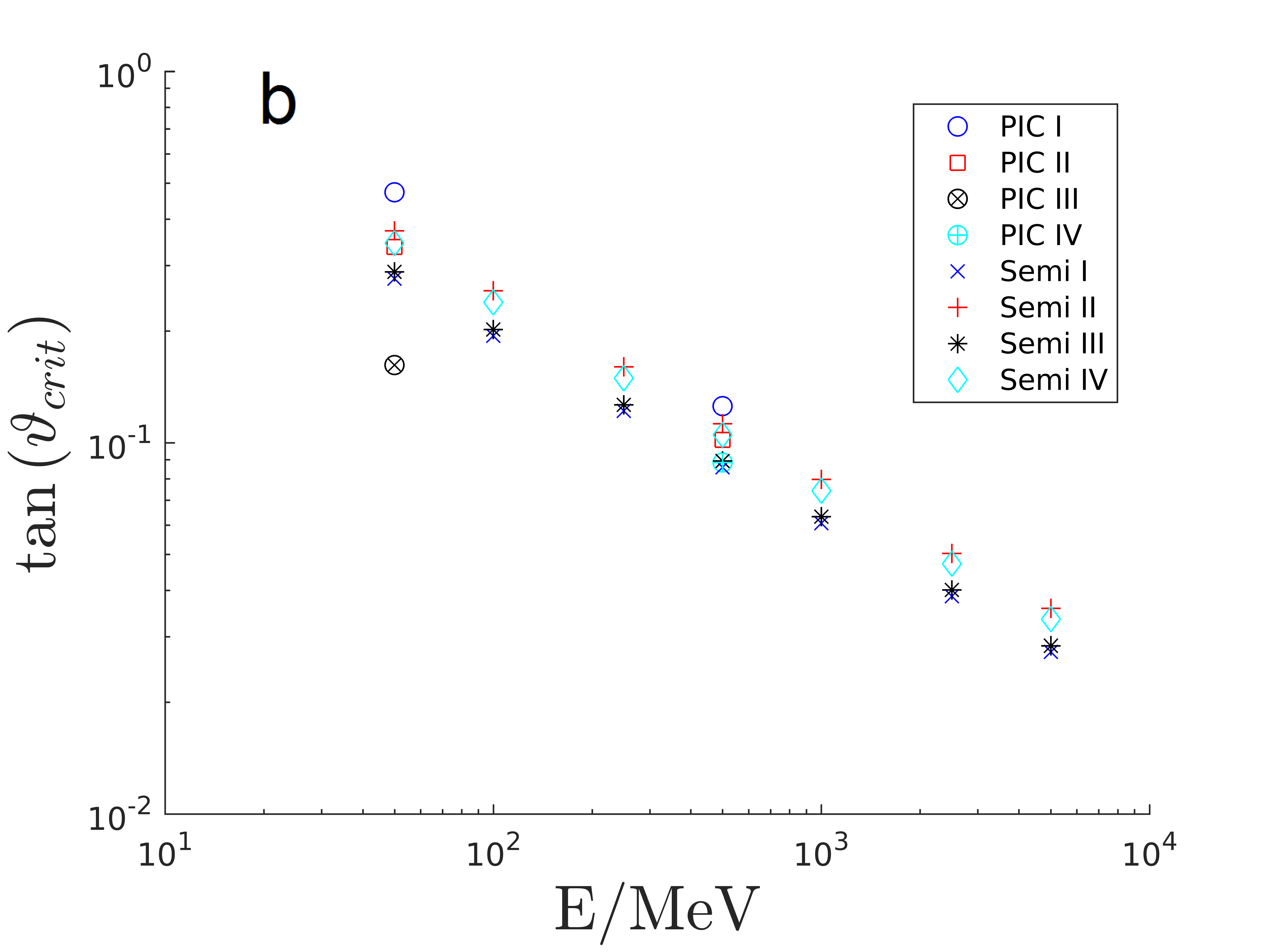}
	\caption{Comparison of the critical injection angles $\vartheta_\text{crit}$ from PIC simulations, test particle simulations and analytical approximations for (a) homogeneous plasma and (b) channeled plasma  with parabolic density profile $\rho_\text{ion}\propto r^2$. The blue (solid) and red (dashed) lines belong to analytical predictions from formula Eq.(\ref{eq:tancrit}) for position I and II in Fig.\ref{pic:winkel}a. The markers $\times$, $+$, $\ast$ and $\diamond$ are interpolated values for $\tan(\vartheta_\text{crit})$ from simulations of test particles in the semi-analytical model, while the enclosed symbols $\bigcirc$, $\square$, $\otimes$ and $\oplus$ are interpolated values from PIC simulations. The numbers I to IV refer to the initial positions in Fig.\ref{pic:winkel}.}
	\label{pic:theta_vlg}
\end{figure*}

Other PIC simulations were performed for different initial energies of injected electrons, ranging from $50$ MeV to $5$ GeV. The initial witness bunch positions relative to the driver are labeled by the numbered positions in Fig.\ref{pic:winkel}a and \ref{pic:winkel}b. In both simulation series positions I and II are chosen such that the center of mass of the witness bunch is located on the blow-out symmetry axis. For homogeneous plasma (Fig.\ref{pic:winkel}a) the center of mass in position III has the same $\xi$-coordinate as in I and an additional radial shift close to the electron sheath. For the channeled plasma (Fig.\ref{pic:winkel}b) the center of mass in positions III and IV have the same $\xi$-coordinates as in I and II respectively. Similar to simulations for homogeneous plasma both bunches are radially shifted close to the electron sheath.

To compare the critical injection angles observed in PIC simulations to those discussed in section \ref{sec:simu} and to analytical predictions from Eq.\ref{eq:tancrit}, we plot these information into one cumulative diagram. The result for homogeneous background plasma is presented in Fig. \ref{pic:theta_vlg}a whereas Fig. \ref{pic:theta_vlg}b is an evaluation for parabolically channeled plasma. In both diagrams $\tan(\vartheta_\text{crit})$ is plotted versus the initial energy of injected electrons. The blue (solid) line in Fig. \ref{pic:theta_vlg}a belongs to the analytical prediction from formula Eq.(\ref{eq:tancrit}) for position I while the red (dashed) line belongs to the analytical prediction for position II. 
The Semi-markers are interpolated values for $\tan(\vartheta_\text{crit})$ from test particle simulations we discussed in the previous section. A comparison to the analytical predictions shows that those markers representing simulations in homogeneous plasma with initial positions near the symmetry axis cover the predictions for all energies while the other markers representing simulations with initial electron positions near the blow-out border are positioned well below the lines. In contrast to that we observe that the Semi-markers for test particle simulations in channeled plasma are much closer. This circumstance coincides with the discussion of Fig. \ref{pic:winkel}a and \ref{pic:winkel}b in the previous section so that we conclude that injection of pre-accelerated electron bunches into a blow-out in homogeneous plasma should be done preferably near the symmetry axis. In channeled plasma small deviations in radial direction are possible as long as the injected bunch is well focused.

The critical angles we observe in PIC simulations are represented by enclosed markers in Fig. \ref{pic:theta_vlg}a and Fig. \ref{pic:theta_vlg}b. In both figures we clearly see that the $\bigcirc$ markers are close to the analytical predictions and the test particle simulations. This indicates that external injection of pre-accelerated electron bunches on-axis and near the bubble center is a promising method for both homogeneous and channeled plasma. For on-axis injection near the bubble back the $\square$ markers in Fig. \ref{pic:theta_vlg}b show a similar agreement to the simplified bubble model. In contrast to that the $\square$ markers in Fig. \ref{pic:theta_vlg}a are placed well below the predictions of the simplified models but still in the same order. From this we conclude that the trapping process for on-axis injection can be modeled by simplified models quite well and that there is not much difference for this kind of injection between deep plasma channels and homogeneous plasmas.

For an off-axis injection of pre-accelerated electron beams (see PIC II and PIC III markers) our PIC simulations confirm what we already observed: In homogeneous plasma it is favorable to inject on-axis while in channeled plasma with $\rho_\text{ion}\propto r^2$ the distance to the symmetry axis is less important for trapping. However, the possibility to trap bunches which have been injected off-axis in channeled plasma is compensated by the need for a higher initial focussing of the witness beam.

\section{Conclusion}
\label{sec:conclusion}
In this work we study electron side injection into a blow-out in homogeneous and channeled plasma. We discuss the critical injection angle $\vartheta_\text{crit}$ for which at least 90\% of the injected particles are trapped and show that external injection into blow-outs is less critical in deep channels than in homogeneous plasma. A comparison of our results from analytical predictions to test particle simulations in a quasi-static blow-out model and to PIC simulations shows that in homogeneous plasma it is favorable to inject on-axis while in channeled plasma it is possible to trap bunches which have been injected off-axis. However, this advantage is compensated by the need for a higher initial focussing of the injected electron beam.

\begin{acknowledgments}
	This work has been supported in parts by DFG project PU-213 and BMBF project 05K2016.
\end{acknowledgments}

\bibliographystyle{prsty}
%\bibliography{Plasma_Physik} 

\end{document}